\documentclass[USenglish,onecolumn]{article}

\usepackage[numbers]{natbib}
\usepackage[big]{dgruyter}
\usepackage{color}

\newcommand{\nc}{\newcommand}
\nc{\nn}{\nonumber}

\def\e{\mathcal{E}}

\def\ee{\textrm{e-e}}
\def\eph{\textrm{e-ph}}
\def\eimp{\textrm{e-imp}}
\def\epht{\textrm{photon-e}}
\nc{\XYZ }{}

\begin{document}

\articletype{Research Article{\hfill}Open Access}

\author*[1]{Y.~Sivan}
\author[2,3]{I.~W.~Un}
\author[4]{S.~Sarkar}

\affil[1]{School of Electrical and Computer engineering, Ben-Gurion University of the Negev, P.O. Box 653, Israel, 8410501, E-mail: sivanyon@bgu.ac.il. }

\affil[2]{Key Laboratory of Atomic and Subatomic Structure and Quantum Control (Ministry of Education), Guangdong Basic Research Center of Excellence for Structure and Fundamental Interactions of Matter, School of Physics, South China Normal University, Guangzhou 510006, China.}

\affil[3]{Guangdong Provincial Key Laboratory of Quantum Engineering and Quantum Materials, Guangdong-Hong Kong Joint Laboratory of Quantum Matter, South China Normal University, Guangzhou 510006, China.}

\affil[4]{Department of Physics and Nanotechnology, SRM Institute of Science and Technology, Kattankulathur-603 203, India.}

\title{\huge Ballistic vs. diffusive transport in metals}
\runningtitle{Ballistic vs. diffusive transport in metals}

\begin{abstract}
{Using the Boltzmann transport model, we show that, somewhat unintuitively, ballistic transport of electrons in metals is weaker than diffusive transport. This happens because the femtosecond-scale collision rates of the non-thermal electrons makes their mean-free path negligible. Our predictions are correlated with various photoluminescence and nonlinear optics experimental examples both for Continuous Wave (CW) and pulsed illumination, and open the way to easy modelling of the non-thermal electron distributions in metal nanostructures of arbitrary complexity. }
\end{abstract}

\keywords{Electron non-equilibrium, plasmonics, transparent conducting oxides, metal nanostructures, photoluminescence.}

\journalname{Nanophotonics}

\startpage{1}

\maketitle

\section{Introduction}

Particle and charge transport are fundamental aspects of solid state physics. They also have practical importance, especially in metals, which are part of the backbone of our communications and computing technologies. The basic characteristics of transport in metals were established already in the early days of solid state research. Its modelling usually relied on the simple Drude model, which accounts for (stationary) transport of the electron subsystem assuming it is at low temperatures and in thermal equilibrium~\cite{Ashcroft-Mermin}. Under such conditions, the transport is diffusive
. If the electrons are subject to pulsed illumination, then, one has to distinguish between the electron and phonon subsystems~\cite{Two_temp_model,Stoll_review}. The model used to describe their dynamics, the so-called Two Temperature Model (TTM), shows that the much smaller heat capacity of the electrons causes them to experience stronger heating. Electron-phonon interactions cause the temperatures of the two subsystems to equalize within a few picoseconds; 
they also lock their spatial dynamics (see e.g.,~\cite{Maznev-2011,Thesis_Block,Jaime_2021}). Due to the short temporal and spatial scales associated with the electron transport, an experimental demonstration of the differences between the electron and phonon diffusive transport (in the direction perpendicular to the illumination direction) was obtained only recently~\cite{ICFO_Sivan_metal_diffusion,Gao_negative_diffusion_2023,Tielrooij_Sivan_negative_diffusion,Giri_Hopkins_diffusion}; the observed non-trivial spatio-temporal dynamics was captured quantitatively within the TTM (see e.g.,~\cite{Rotenberg_PRB_09,Sivan_Spector_metal_diffusion,Spector_Derevyanko_2022,Abajo_DTU_PL}).


Only a few studies attempted to describe the electron transport beyond the above description, namely, the ballistic transport that precedes it and occurs during the post-excitation and pre-scattering stage of the electrons. 
Indeed, the experimental studies of lateral transport mentioned above were careful not to discuss the ballistic stage\footnote{An exception are the claims in~\cite{Giri_Hopkins_diffusion} which were nevertheless based on an experimental system whose spatial and temporal resolution were no different from those in earlier work; the likeness to earlier results is apparent. }. A somewhat refined model adopted the extension of the TTM which includes the dynamics of the total non-thermal energy, and included also the possibility of its diffusive expansion~\cite{Sivan_Spector_metal_diffusion}; this approach led to the prediction of fs-scale decay of heat and permittivity gratings in metal layers. In the context of longitudinal transport through a thick metal layer, there have been several studies of ultrafast heat transport (e.g. based on a configuration in which the pump and probe are incident on opposing interfaces of a metal film, see~\cite{Brorson1987,optic_excitation}). Theoretical modelling of this effect accounted for the transport via a 1D diffusion term, and used an augmented penetration depth to account phenomenologically for ballistic transport~\cite{optic_excitation,Ballistic_e_diffusion}. Budai {\em et al.}~\cite{Dombi_Nat_comm} studied the same configuration with Continuous Wave (CW) light. They also concluded that the non-thermal electrons do not penetrate the layer thickness.



However, it was implicit from the above studies that the actual electron transport may be faster once ballistic transport is accounted for properly. To do that, one needs to abandon the macroscopic TTM approach in favour of a microscopic model which accounts for the dynamics on the electron energy level. 
While Khurgin showed that the non-thermal electrons are effectively generated near the metal interfaces~\cite{Khurgin-Faraday-hot-es}, their consequent spatio-temporal dynamics and/or spatial steady-state distributions were not well studied~\cite{Dombi_Nat_comm}. Indeed, microscopic models for the non-thermal electron dynamics such as the Boltzmann equation (BE) almost always relied on a local model (e.g.,~\cite{Stoll_review,Riffe-Wilson-2022}), and the few studies that went beyond the BE to a fully quantum description (e.g.,~\cite{Govorov_ACS_phot_2017,GdA_hot_es,Brown_PRL_2017}, or more advanced Density functional theory (DFT) -based calculations, e.g.~\cite{Lischner_hot_es,Lischner_2023}) did not discuss spatio-temporal gradients and currents of field and/or charge nor their dynamics in detail. However, a-priori there is no reason to rule out such dynamics in light of the fs-scale electron collision mechanisms and the well known diffusive expansion. In particular, as shown in~\cite{Dubi-Sivan,Dubi-Sivan-Faraday}, the steady-state distribution may differ significantly from the non-thermal generation spectra even under uniform CW illumination.

Several attempts to model the transport of non-thermal electrons stand out in the literature. First, Tien {\em et al.}~\cite{Tien_1993} employed the Boltzmann Transport Equation (BTE), namely, the local Boltzmann equation augmented by the spatial part of the hydrodynamic derivative of the electron distribution. However, the model of~\cite{Tien_1993} included a classical excitation term and a near-equilibrium assumption in which the deviation from equilibrium is ignored, such that this study focussed only on the (effective) electron temperature dynamics and spatial distribution. 
Jermyn {\em et al.}~\cite{NESSE} improved upon~\cite{Tien_1993} by replacing the classical approach for the photon absorption with DFT calculations with a hydrodynamic description of the transport within the BTE. 
However, while this work described some aspects of the transport undergone by thermal electrons in great detail (e.g., the detailed absoprtion mechanisms and tracking the number of collisions events for each electron etc.), it did not treat the non-thermal electrons, i.e., those electrons with a high excess energy compared to the Fermi energy which cannot be described by a the Fermi-Dirac function. 


Another attempt to model the transport of non-thermal charge carriers with the BTE was made by Rudenko {\em et al.} in~\cite{Moloney_PRB_2021}; it 
studied self-consistently the field, electron distribution and permittivity in a small sphere. However, like~\cite{Tien_1993}, it focussed on the spatio-temporal dynamics of the (effective) electron temperature rather than on the dynamics of the non-thermal electrons.



In this work, we go beyond the above work and provide a comprehensive study of the transport of electrons in a Drude metal, including both thermal and non-thermal electrons during both the ballistic and diffusive stages. In Section~\ref{sec:formulation} we describe our formulation. Briefly, we compute the electron collision rates in momentum space and convert the resulting expressions to energy space; the excitation is computed directly in energy space, without detailing the specific mechanism as in~\cite{hot_es_Atwater,Abajo_Tagliabue_PL}; finally, the transport is described on the hydrodynamic level via the BTE, such that this aspect is the only one that is not given a full quantum mechanical treatment. This constitutes the simplest model suitable to model non-thermal electron transport.

In Section~\ref{sub:CW_illum}, we consider the simple configuration of a thin film of Drude metal illuminated by a CW beam. To distinguish the current study further from earlier work, we focus on Indium-Tin oxide (ITO) as the plasmonic material. First, we reproduce the thermal aspects of the problem, and show that they are accompanied by an essentially negligible spatial dynamics of the non-thermal electrons. Then, using the simple analytic solution for the steady-state electron non-equilibrium of~\cite{Dubi-Sivan-Faraday}, we show that this behaviour can be explained by the very short mean free path (MFP) of these electrons, originating from their frequent collisions. In Section~\ref{sub:pulsed_illum} we show the analogous results for pulsed illumination. We again show that the transport in the ballistic stage is weaker compared to the diffusive stage and identify the signature of the electron-electron collision rate on the spatial broadening rate. Finally, Section~\ref{sec:exps} reviews several experiments that are in line with our theoretical predictions, and Section~\ref{sec:discussion} summarizes the implications of our work and provides an outlook. Primarily, we conclude that in most cases, one can determine the non-thermal electron distribution without solving the BTE nor the BE at all; it suffices simply to compute the electric field and temperature. Considering that non-thermal electron research has suffered from poor understanding of the electron distribution due to the complex momentum space modelling (e.g.,~\cite{Manjavacas_Nordlander,Govorov_ACS_phot_2017,hot_es_Atwater,Lischner_hot_es,Riffe-Wilson-2022}), our approach has the potential to significantly promote the comprehension of electron non-equilibrium within the community.

\section{Formulation}\label{sec:formulation}
We adopt the energy-space Boltzmann formulation described in~\cite{Dubi-Sivan,Un-Sarkar-Sivan-LEDD-I,Un-Sarkar-Sivan-LEDD-II} for the non-equilibrium electron distribution function $f$; it depends on the electron energy $\e$, on time $t$ as well as on the position $r$, and determined by electron-electron ($e-e$), electron-phonon ($e-ph$), photon-electron ($photon-e$) and electron-impurity ($e-imp$) interactions. This approach was shown to be suitable for metal nanostructures whose features are no smaller than a few nm~\cite{Stoll_review,Govorov_ACS_phot_2017,Khurgin-Levy-ACS-Photonics-2020,Sarkar-Un-Sivan-DM} and for pulses no shorter than a few femtosecond long. Here, however, we add the spatial part of the hydrodynamic derivative~\cite{Boardman-book}, namely,
\begin{align}\label{eq:BE_E}
\dfrac{\partial f(\e,r,t)}{\partial t} + \left(\dfrac{\partial f}{\partial t}\right)_{trans} = \left(\dfrac{\partial f}{\partial t}\right)_\ee + \left(\dfrac{\partial f}{\partial t}\right)_\eph + \left(\dfrac{\partial f}{\partial t}\right)_\epht + \left(\dfrac{\partial f}{\partial t}\right)_{e-imp}.
\end{align}
The formulation of the various terms has been described previously (e.g.,~\cite{Dubi-Sivan,Dubi-Sivan-Faraday}), and will not be repeated here. Note only that while we account for momentum conservation in the calculation of $\ee$ and $\eph$ interactions\footnote{Note, however, that momentum conservation in $\eph$ interactions is not really necessary for noble metals~\cite{Un-Sarkar-Sivan-LEDD-I}. }, as in~\cite{delFatti_nonequilib_2000,Dubi-Sivan}, our treatment of the photon-e interactions ignores momentum conservation~\cite{Sarkar-Un-Sivan-DM} (i.e., it assumed energy-independent matrix elements); this assumption limits the quantitative accuracy of our results, which are nevertheless qualitatively correct, as seen is several recent experimental measurements~\cite{Dubi-Sivan-MJs,Gabelli,Kumagai-ACS-phot-2023}\footnote{Having said that, we note that unlike in~\cite{hot_es_Atwater} where the energy states were assumed to have a constant width, and unlike~\cite{GdA_hot_es} which avoid broadening the energy states, but like in~\cite{Lischner_hot_es,Un-Sarkar-Sivan-LEDD-I}, we compute the energy state broadening in a consistent manner.}$^,$\footnote{Also note that as in~\cite{delFatti_nonequilib_2000}, we do not include the frequency-dependence of the photon-e interaction term 
as done in~\cite{GdA_hot_es}, since we deal primarily with CW; under the conditions of the pulses we do simulate below, these effects are minor. }. 
We also emphasize that the $\ee$ interaction term is computed without the relaxation time approximation (RTA); this approach is more accurate, but mainly unavoidable, because unlike uniformly illuminated electron systems, it is not obvious how to pre-determine the spatial temperature profile necessary for using the RTA
. Instead, we adopt a Thomas-Fermi type interaction kernel, as in~\cite{delFatti_nonequilib_2000,Un-Sarkar-Sivan-LEDD-I,Un-Sarkar-Sivan-LEDD-II}.

The only difference to previous implementations is that here we allow the electric field profile to be non-uniform; this gives rise, naturally, also to charge gradients and currents (i.e., transport). The transport term is implemented in this context for the first time here; it equals $\vec{v}(\e) \cdot \nabla f$ where $\vec{v}(\e) = \partial \e(\vec{k}) / \hbar \partial \vec{k} / m_e^*$ is the electron velocity, with $\e(\vec{k})$ being the electron dispersion relation. This constitutes the Boltzmann Transport equation (BTE). Such hydrodynamic approach is popular and successful for modelling linear and nonlinear nonlocal optical effects in metals (e.g.,~\cite{Boardman-book,scalora_model_SHG_THG,DTU_nonlocal,Ciraci:2012_Science,Ciraci:2012_PRBR,Lienau_hydrodynamic,Scalora-ITO-2020}); it provide a good compromise between accuracy and simplicity.


As in~\cite{Dubi-Sivan,Dubi-Sivan-Faraday}, we complement the Boltzmann formulation with an equation for the total energy of the phonons (or equivalently, for the phonon temperature). This enables ensuring that the photon energy that is added to the electron subsystem upon absorption is balanced by energy transfer from the phonons to the environment. As shown in~\cite{Dubi-Sivan,Dubi-Sivan-Faraday}, this is essential in order to get quantitatively correct results for the electron distribution near the Fermi energy. Since on one hand the electron and phonon temperatures are similar under CW illumination, and since we are interested only in the early stages of the dynamics following a short pulse, we neglect phonon diffusion. 

We emphasize that our approach combines elements of quantum mechanics and classical physics. Indeed, our approach for the photon-electron interactions reproduces the photoelectric effect and the electron collisions are treated no differently than in more advanced models~\cite{hot_es_Atwater,GdA_hot_es}. Yet, the spatial part of the hydrodynamic derivative refers to the electrons as localized entities, i.e., as particles rather than as waves. Nevertheless, below we show that this model not only has the advantage of relative simplicity over full-scale quantum mechanical models, it is also capable of explaining various experimental results, and offers a very simple approach to understand and model quantitatively the deviation of the electron distribution from thermal equilibrium in complex nanostructures.


\section{Results}\label{sec:results}

\subsection{CW}\label{sub:CW_illum}
First, we study the non-equilibrium electron distribution under CW illumination. We consider a few nm thick ITO layer illuminated by a cylindrically symmetric focused Gaussian beam at normal incidence (e.g.,~\cite{Abajo_Tagliabue_PL}); specifically, $E = E_0 e^{- (\rho / \rho_0)^2 - i \omega_0 t}$, $\rho_0$ being the illuminating beam radius (henceforth, the waist). In that regard, we assume that the illumination is uniform across the layer thickness, and consider the charge and heat transport only in the radial direction. We employ a perfectly matched layer (PML) at the outer edges of the domain in order to truncate the transverse size of the computational domain, see Appendix~\ref{app:PML}. We chose $\lambda_0 = 2 \pi c / \omega_0 = 1300$nm, illumination intensity of $100 KW/cm^2$
, and waist of $\rho_0 = \lambda$. The material parameters are taken from~\cite{Un-Sarkar-Sivan-LEDD-I,Un-Sarkar-Sivan-LEDD-II}. In practice, the steady-state is reached by gradually turning on the electric field, until convergence. 

\begin{figure}[h]
\centering
\includegraphics[width=1\textwidth]{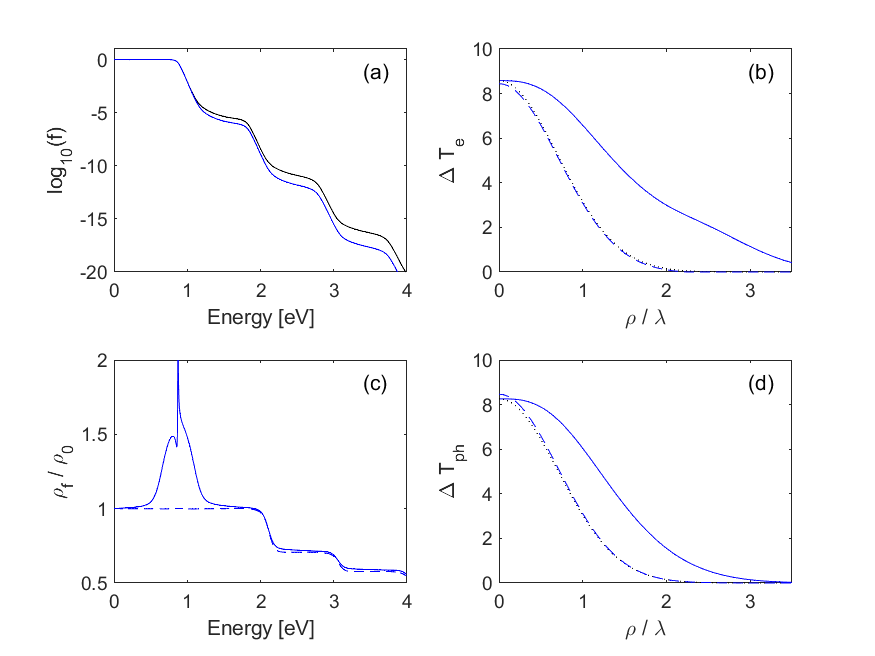}
\caption{(Color online) (a) The steady-state non-equilibrium electron distribution in a thin ITO layer under CW illumination; the black (blue) lines are taken at the beam center (waist). The illuminating beam has (spatial peak) intensity of $100 KW / cm^2$ and central wavelength of $1300$nm. (b) The extracted electron temperature profile (blue); it is much wider than the spatial profile of the extracted temperature in the absence of the transport term (dashed blue line); it is indistinguishable from the illuminating beam profile (dotted black line). (c) The spatial width of the non-equilibrium electron distribution (normalized by the illuminating beam size) as a function of the electron energy. The dashed blue line shows the width in the absence of the transport term. (d) The spatial profile of the phonon temperature. The dashed blue line shows the phonon temperature in the absence of transport; it is indistinguishable from the illuminating beam profile (dotted black line). }
\label{fig:widths}
\end{figure}

Fig.~\ref{fig:widths}(a) shows the electron distribution at the beam center and at its spatial waist $\rho_0$. One can observe the step structure seen already for uniformly illuminated systems (e.g.,~\cite{non_eq_model_Rethfeld,Italians_hot_es,Dubi-Sivan,Dubi-Sivan-Faraday,Kumagai-ACS-phot-2023}); the height of the non-thermal electron shoulders is lower at the waist due to the weaker illumination at this position. Fig.~\ref{fig:widths}(b) shows the spatial distribution of the extracted effective electron temperature $T_e$ (based on a computation of the Fermi energy at each point in space, and matching to the thermal distribution with the same total electron energy); it exhibits the expected (diffusive) broadening compared to the illumination spot that one obtains from solving the classical heat equation (i.e., single temperature model). This broadening naturally originates from the inclusion of the transport term. 
These are the expected (thermal equilibrium) features of the problem.

Here, however, we would like to go beyond the thermal aspects, and consider also the transport of the non-thermal part of the electron non-equilibrium distribution. To do that, we compute the spatial width of the (steady-state) electron distribution as a function of the electron energy, namely,
\begin{equation} \label{eq:rho_f}
\rho_f(\e) = \left|\int \rho^2 \left[f(\e,\rho) - f^T(\e;T_e = 300K)\right] d^2\rho\ \Big/ \int \left[f(\e,\rho) - f^T(\e;T_e = 300K)\right] d^2\rho\right|,
\end{equation}
where $f^T$ is the thermal, i.e., Fermi-Dirac distribution. 
Fig.~\ref{fig:widths}(c) indeed shows a broadening near the Fermi energy compared to the illumination spot, 
inline with the broadening of the electron temperature spot. Peculiarly, the broadening shows some asymmetry with respect to the Fermi energy. 
On the other hand, the spatial width of the non-thermal electrons, those with high excess energy above the Fermi energy, is not much different from that of the illumination spot. 
In that sense, the non-thermal electrons do not exhibit significant transport, or equivalently, the ballistic transport effects are so weak so that they can simply be ignored and their spatial distribution is given by the electric field distribution. This somewhat non-intuitive observation is the main result of this work.

\begin{figure}[h]
\centering
\includegraphics[width=1\textwidth]{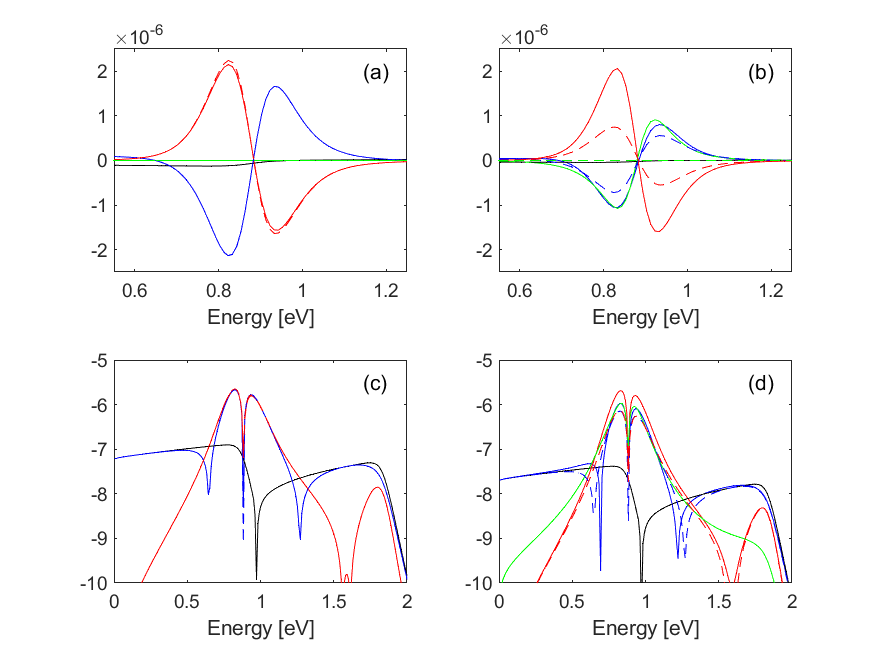}
\caption{(Color online) The $\epht$ (black), $\ee$ (blue), $\eph$ (red) and transport terms (green) as a function of the electron energy. The top (bottom) row shows the values on a linear ($log_{10}$) scale, whereas the left (right) column shows the results at the beam center (spatial waist $\rho_0$). The dashed lines show the various terms when the transport term is set to zero. }
\label{fig:derivatives}
\end{figure}

Insight into the underlying physics of this result can be obtained by comparing the magnitudes of the various terms in the BTE~(\ref{eq:BE_E}). Fig.~\ref{fig:derivatives} shows that while in the absence of transport (as well as at the beam center) the distribution near the Fermi energy is determined by the balance of $\ee$ and $\eph$ collisions (see Figs.~\ref{fig:derivatives}(a) and~(c)~\footnote{This is what happens under uniform illumination~\cite{Dubi-Sivan-Faraday}. }), when the spatial gradients are non-zero (in particular, away from the beam center), the magnitude of the transport term is comparable to both electron collision terms (see Figs.~\ref{fig:derivatives}(b) and~(d)). For our particular choice of parameters, the presence of spatial gradients leads to a $3$-fold increase of the magnitude of the $\eph$ term and a slight increase of the $\ee$ term; the transport term turns out to be comparable to the $\ee$ term, such that they balance together the $\eph$ term. To qualitatively understand this behaviour, we recall that near the Fermi energy, the transport term can be approximated by
\begin{equation}
- v(\e) \frac{\partial f}{\partial r} \sim - v(\e) \frac{\partial f^T}{\partial T_e} \frac{\partial T_e}{\partial r} \sim - v(\e) \frac{\e - \mu}{T_e} \frac{\partial f^T}{\partial \e} \frac{\partial T_e}{\partial r}.
\end{equation}
Since the derivative of the Fermi-Dirac function with respect to the energy is negative, and since the spatial derivative of the electron temperature is negative as well, the transport term is negative below the chemical potential, and positive above it. In that sense, it behaves like the $\ee$ term, see Fig.~\ref{fig:derivatives}.

These results indicate that not only the transport term causes the expected broader spread of the electron energy in space, but it also causes a larger amount of energy to be transferred from the electron subsystem to the phonon subsystem. Indeed, our simulations (Fig.~\ref{fig:widths}(d)) show that the inclusion of the transport term causes the phonon temperature spot to broaden
, and the overall phonon energy (integrated across the phonon temperature spot) to increase by $\sim 25\%$. 

\begin{figure}[h]
\centering
\includegraphics[width=1\textwidth]{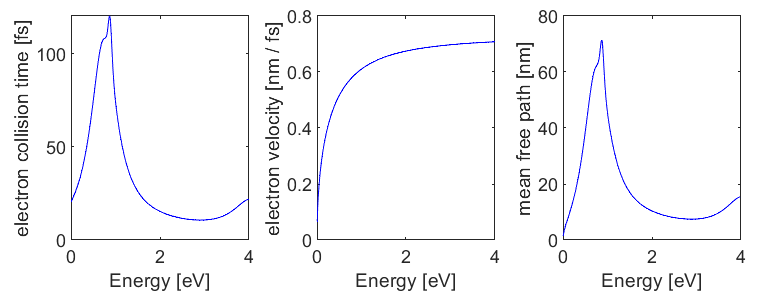}
\caption{(Color online) (a) The electron collision time, derived from the functional derivatives of the electron scattering integrals~\cite{Un-Sarkar-Sivan-LEDD-I} and Matthiessen's rule, (b) the electron velocity and (c) their product, i.e., the mean free path, all shown as a function of the electron energy. These results are analogous to~\cite[Fig.~5]{hot_es_Atwater}. }
\label{fig:MFP}
\end{figure}

Fig.~\ref{fig:derivatives} also shows that on the other hand, at energies far from the Fermi energy, the $\eph$ and transport terms decrease rapidly with energy, such that the electron distribution is determined by the balance of photon absorption and $\ee$ interactions (again, as for uniformly illuminated systems, see~\cite{Dubi-Sivan-Faraday}). In fact, using the analytic solution for the steady-state non-equilibrium electron distribution attained under these conditions (and within the RTA and Fermi liquid theory~\cite{Quantum-Liquid-Coleman}) motivates the observation of negligible transport exhibited by the non-thermal electrons. Indeed, the deviation from equilibrium is~\cite{Dubi-Sivan,Dubi-Sivan-Faraday}
\begin{equation}
f(\e) - f^T(\e,T_e) \sim \delta_E f^T(\e - \hbar \omega,T_e).
\end{equation}
Thus, the transport term can be approximated by $\sim v(\e) f(\e) / \rho_0 \sim v(\e) \delta_E f^T(\e - \hbar \omega) / \rho_0 \sim v(\e) \tau_e(\e) / \rho_0$ ($\delta_E \sim \tau_e |E|^2$ being the height of the non-thermal electron distribution, see derivation in~\cite{Dubi-Sivan-Faraday}), so that it scales with the mean-free-path of the non-thermal electrons. As shown in Fig.~\ref{fig:MFP}, the mean free path decreases rapidly away from the Fermi energy. This somewhat classical insight was already obtained in~\cite{hot_es_Atwater}.







\subsection{Pulsed illumination}\label{sub:pulsed_illum}
We now study the spatial dynamics of the electrons in the same ITO layer following illumination by a short pulse. In particular, we now set the illuminating electric field to be
\begin{equation}
E = E_0 e^{ - 2 log(2) (t / \tau_p)^2 - (\rho / \rho_0)^2 - i \omega_0 t},
\end{equation}
where the peak intensity is now set to 500 $KW/cm^2$ and $\tau_p = 100$fs.

\begin{figure}[h]
\centering
\includegraphics[width=1\textwidth]{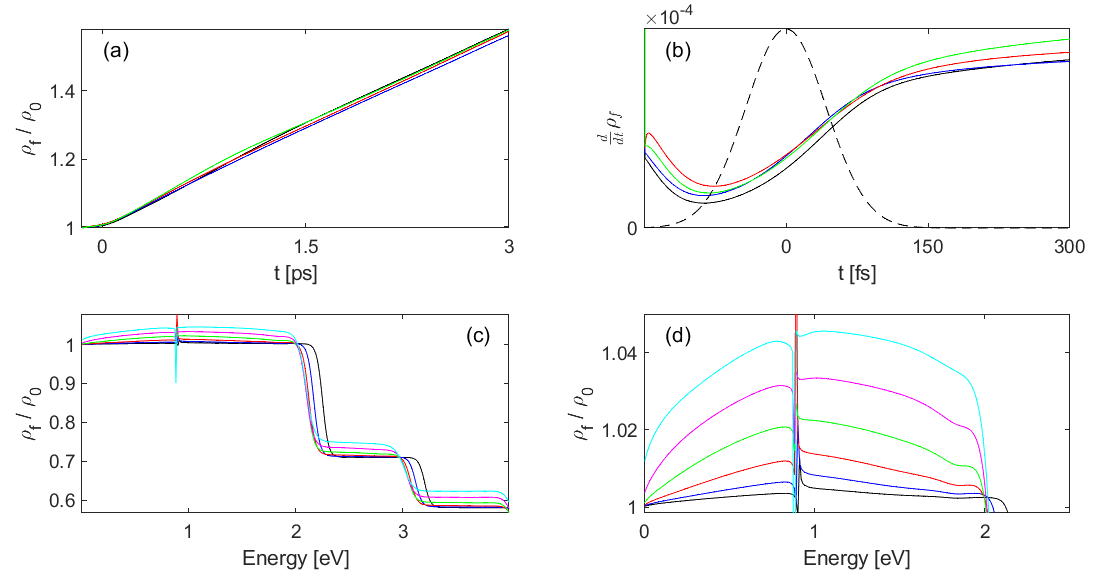}
\caption{(Color online) (a) The width of the electron distribution $f$ as a function of time for $\e = 0.6$ eV (black), $\e = 0.8$ eV (blue), $\e = 1$ eV (red) and $\e = 1.2$ eV (green). (b) The time derivative of the curves in (a), zoomed-in to the early stages of the dynamics; the dashed black line represents the temporal shape of the illumination. (c) The width of the electron distribution $f$ as a function of electron energy for $t = -120, -90, -60, -30, 0, 30$ fs (in black, blue, red, green, magenta and cyan, respectively). (d) Same as (c) for a narrower energy range close to the Fermi energy. }
\label{fig:pulses}
\end{figure}

Again considering the width of the electron distribution $\rho_f$~(\ref{eq:rho_f}), we now observe in Fig.~\ref{fig:pulses} monotonic expansion of the spot size 
at a rate which is, somewhat surprisingly, weakly dependent on the electron energy (at least up to $\e \sim \e_f + \hbar \omega$)\footnote{The higher energy electrons follow the same step structure as for the CW case, Fig.~\ref{fig:widths}(c) and similar dynamics as those of the thermal electrons. }. 
Further, the early stages of the dynamics (Fig.~\ref{fig:pulses}(b)) show a relatively slower expansion compared to the later diffusive stage; the effect is similar for different initial spot sizes (not shown). This behaviour may seem to be at odds with the linear and square root (i.e., {\em slower}) growth rates, respectively, usually associated with the ballistic and diffusive regimes. 
However, this behaviour can be understood as another manifestation of the shortness of the MFP of non-thermal electrons. In that sense, initially the pulse generates a relatively large number of non-thermal electrons with a large excess energy (above the Fermi level), however, these electrons are relatively immobile. As the electron subsystems thermalization proceeds, the high excess energy non-thermal electrons lose their excess energy and get converted into thermal electrons, which are more mobile; eventually, the systems settles into a constant spot growth rate (late stages seen in Fig.~\ref{fig:pulses}(a)). In that regard, a more careful look at Fig.~\ref{fig:pulses}(d) reveals that the electrons near the Fermi energy exhibit a somewhat stronger expansion; this is inline with the energy dependence of the MFP, which is dominated by $e-e$ collisions for the non-thermal electrons (Fig.~\ref{fig:MFP}(c)). 







\section{Experimental evidence}\label{sec:exps}
In recent years, several classes of measurements were performed in order to identify the signature of non-thermal electrons. While contrary to earlier claims, photocatalysis experiments were shown to be dominated almost always by thermal effects (see, e.g.,~\cite{Dubi-Sivan-APL-Perspective,Baffou-Quidant-Baldi,Jain_viewpoint,anti-Halas-NatCat-paper}), there have been several convincing demonstrations of non-thermal electrons in molecular junctions (see discussion in~\cite{Dubi-Sivan-MJs}), tunnel junctions~\cite{Gabelli,Kumagai-ACS-phot-2023} as well as photoluminescence (PL) experiments. As the latter class offers the cleanest signature of the non-thermal electrons for weak illumination~\cite{Sivan_tPL}, we now focus on PL experiments where signatures of the weak ballistic transport predicted above were observed.

First, the observation of the negligible transport experienced by non-thermal electrons is inline with the measurements of the area of an illuminated Au film from which photoluminescence (PL) emerges~\cite{Abajo_Tagliabue_PL}. Indeed, this work showed experimentally that this area is essentially similar to the illuminated area. Since CW PL originates from the non-thermal electrons~\cite{Sivan-Dubi-PL_I}, this finding implies that non-thermal electrons do not undergo any significant (ballistic) transport. Our predictions strengthen the conclusion drawn in the original work which was based on a calculation that neglects transport altogether.



In this respect, our calculations give rise to a prediction that could be verified in a (indeed challenging) future experiment. Specifically, as seen in Fig.~\ref{fig:widths}(c), the spatial width of the higher energy state occupation (i.e., those electrons with $\e > \e_F + \hbar \omega$) is even narrower than the illumination waist $\rho_0$. Indeed, the electron distribution extends as $e^{- (\rho / \rho_0)^4}$ and $e^{- (\rho / \rho_0)^6}$ for energies in the second and third non-thermal electron ``shoulders'', i.e., the width of the energy distribution is $\sim \rho_0/\sqrt{2}$, $\sim \rho_0/\sqrt{3}$, respectively. In PL measurements, this narrowing would be hard to observe with CW illumination since it involves frequencies deep in the anti-Stokes regime, where the signal is weak. However, as for the electric field dependence of the PL from metals~\cite{Sivan_tPL}, it may be observable with strong pulsed illumination. 

Similarly, when the PL is dominated by interband emission (i.e., recombination of conduction band non-thermal electrons with non-thermal holes in the valence band; see e.g., the elaborate discussions in~\cite{japanese_Mg_PL_2023,Abajo_Tagliabue_PL}), the width of the PL spot is primarily determined by the large number of non-thermal holes created by interband absorption (see~\cite[Fig.~S10b]{Abajo_Tagliabue_PL}). Since those non-thermal holes exhibit a very short mean free path too, then, despite the wider distribution of the electrons in the conduction band that undergo the recombination, the width of the interband PL spot is also expected to be field-limited. This observation a-posteriori justifies the PL-based imaging used in, e.g.,~\cite{Bouhelier_PRL_2005,FJ_Luis_SHG,Quidant_modes,Xu_NTU_2018,Imura_2024} 
which all relied on interband-based PL in Au nanostructures; in that sense, our results imply that this imaging technique would provide lower resolution when it is based on the thermal-like emission that dominates the PL for illumination with strong short pulses (see discussion in~\cite{Sivan_tPL}).

\section{Discussion and Outlook}\label{sec:discussion}
We have provided a formulation for the calculation of the non-uniform non-equilibrium electron distribution in illuminated metal nanostructures which bridges the microscopic and macroscopic pictures for both the ballistic and diffusive stages of electron transport.

We showed that, somewhat unintuitively, the ballistic transport stage is much weaker than the diffusive stage, so that the different positions in space are effectively decoupled, as far as the deviation from equilibrium is concerned. Thus, as long as the inhomogeneity of the illumination is much longer than the mean free path of the non-thermal electrons, the (steady-state) temperature of metal nanostructures can be determined using a standard heat equation, and the deviation from thermal equilibrium requires just the knowledge of the electric field distribution. This makes the computation of the non-thermal electron distribution in space nearly trivial, and in fact, means that the inclusion of the spatial part of the hydrodynamic derivative to the formulation, or using the BE altogether, is mostly redundant under these conditions. In that respect, one can simply use the analytic solution for the non-equilibrium distribution obtained in~\cite{Dubi-Sivan-Faraday} for continuously and uniformly illuminated systems also for non-uniformly illuminated structures. This solution serves also to explain the PL under CW illumination, as well as under pulsed illumination (as in~\cite{Sivan_tPL}). Further, a nonthermal model that includes (``just'') diffusive transport (e.g., as in~\cite{Sivan_Spector_metal_diffusion}) is sufficient to capture the electron temperature distribution and dynamics correctly.

In that respect, while for the configuration we studied (infinitely thin layers) the ballistic transport is negligible, it could be significant when illuminating layers whose thickness exceeds the optical penetration depth. Indeed, in such a configuration, our calculations predict that the illuminated facet will enjoy a relatively high density of non-thermal electrons, whereas the regions deeper in the metal will be characterized by a thermal distribution; The region consisting of the non-thermal electrons will be determined by the combination of the skin depth and the (now not much smaller) MFP of the non-thermal electrons. This is indeed in line with the results and modelling in~\cite{Dombi_Nat_comm} of a constantly illuminated thick layer, as well as studies of ballistic transport due to pulsed illumination~\cite{Brorson1987,optic_excitation,Planken_buried_gratings_2018}. On the other hand, our predictions provide a independent explanation to the observation of the absence of significant ballistic transport in Au-Ni~\cite{Bargheer_Au_Ni} and Cu-Pt-Ni~\cite{Bargheer_Cu_Pt_Ni} heterostructures. 

While the predictions of our model are shown to match several experiments, more subtle effects may require a comparison with full quantum mechanical momentum space modelling (e.g.,~\cite{Riffe-Wilson-2022}) or even DFT-based calculation of the matrix elements (e.g.~\cite{Manjavacas_Nordlander,Lischner_hot_es}); these approaches are, unfortunately, far more demanding computationally, as they requires handling the three spatial dimensions separately, and all possible interactions between the associated momentum components. Other improvements of our model may involve the quantum hydrodynamic approach of~\cite{Ciraci_PRX_2021} which includes also quantum pressure, exchange-correlation interactions, the Lorentz force etc. 
and/or going beyond the thermal lattice assumption (e.g., as in~\cite{Baranov_Kabanov_2014,non-thermal_lattice_model_PRX_2016}). Finally, our approach may also be the basis to extensions of models as in~\cite{Shalaev_LID2} for interband absorption associated transport, 
for plasmon-induced drag~\cite{Durach-book-chapter}, Fizeau drag~\cite{Bliokh_Fizeau}, and for the photo-Dember effect~\cite{photo_Dember_Reklaitis}.

\bigskip

{\bf Acknowledgements.} Y.S. thanks A. Yochelis 
for useful discussions. Y.S. was partially funded by a Lower-Saxony - Israel collaboration grant no. 76251-99-7/20 (ZN 3637) as well as an Israel Science Foundation (ISF) grant (340/2020). I.W.U. was funded by the Guangdong Natural Science Foundation.

\appendix











\section{Numerical implementation}\label{app:PML}
The details of the various terms in the  BTE~(\ref{eq:BE_E}) were detailed in various earlier work, see~\cite{Un-Sarkar-Sivan-LEDD-I} specifically for ITO. The current work, however, requires extending that formulation such that it can handle the spatially non-uniform illumination. While the non-uniformity is trivially inherent to the $\epht$ term, this requires, in addition, an approach to truncate the spatial domain. We do that using a Perfectly Matched Layer (PML), as described below.

First, for simplicity of notations, we define
\begin{equation}
g\left[f(\e,\rho,t)\right] = \left(\dfrac{\partial f(\e,\rho,t)}{\partial t}\right)_\ee + \left(\dfrac{\partial f(\e,\rho,t)}{\partial t}\right)_\eph + \left(\dfrac{\partial f(\e,\rho,t)}{\partial t}\right)_\epht + \left(\dfrac{\partial f(\e,\rho,t)}{\partial t}\right)_\eimp, \nn
\end{equation}
and add to the RHS a term which is essentially zero, namely, $\partial f^T / \partial t$, where $f^T$ is the Fermi-Dirac distribution at room temperature (i.e., a time-independent term). We then Fourier transform Eq.~(\ref{eq:BE_E}), introduce the PML transformation~\cite{PML} $\partial / \partial r \to \left(1 + i \sigma(\rho)^{-1} / \omega\right) \partial / \partial r$, and multiply by $\left(1 + i \sigma(\rho) / \omega\right)$. This yields
\begin{align}\label{eq:BE_E2}
\left(- i \omega + \sigma(\rho)\right) \left[\hat{f}(\e,\rho,\omega) - f^T(\e,T_e = 300K)\right) = \left(1 + i \frac{\sigma(\rho)}{\omega}\right) \hat{g}[f(\e,\rho,t)] - v(\e) \frac{\partial \hat{f}(\e,\rho,\omega)}{\partial r}. \nn
\end{align}
We now define $\hat{\psi} = i \frac{\sigma(\rho)}{\omega} \hat{g}[f(\e,\rho,t)]$, and then, after the inverse transform to the time domain, we get
\begin{eqnarray}\label{eq:BE_E3}
\frac{\partial f(\e,\rho,t)}{\partial t} &=& - \sigma(\rho) \left[f(\e,\rho,t) - f^T(\e,T_e = 300K)\right] + g[f(\e,\rho,t)] + \psi(\e,\rho,t) - v(\e) \frac{\partial f(\e,\rho,t)}{\partial \rho}, \nn \\
\frac{\partial \psi(\e,\rho,t)}{\partial t} &=& \sigma(\rho) g[f(\e,\rho,t)]. \nn
\end{eqnarray}
Specifically, we set $\sigma(\rho < \rho_{PML}) = 0$ and $\sigma(\rho \ge \rho_{PML}) = (\rho - \rho_{PML})^2 / \rho_{PML}^2$. The left side of the computational domain is complemented by a Neuman boundary condition, to impose cylindrical symmetry.




\bibliographystyle{unsrt}

\end{document}